\documentclass[manuscript]{emulateapj}

\usepackage{amsmath}
\usepackage{float}
\usepackage{ifthen}
\usepackage{times}
\usepackage{natbib}
\usepackage{rotating}

\shorttitle{Gas reservoirs and star formation 
       in a forming cluster at $z\backsimeq0.2$}
\shortauthors{Jaff\'e et al.}

\begin{document}

\title{Gas reservoirs and star formation 
       in a forming galaxy cluster at $z\backsimeq0.2$}

\author{Yara L. Jaff\'e$^{1}$, Bianca M. Poggianti$^{1}$, Marc A.~W. Verheijen$^2$, Boris Z. Deshev$^{2,3}$, \and Jacqueline H. van Gorkom$^4$}

\affil{$^1$INAF - Osservatorio Astronomico di Padova, vicolo dell' Osservatorio 5, I-35122 Padova, Italy; \textit{yara.jaffe@oapd.inaf.it}}

\affil{$^2$Kapteyn Astronomical Institute, Landleven 12, 9747 AD Groningen, The Netherlands }

\affil{$^3$Tartu Observatory, T\~oravere, 61602, Estonia, }

\affil{$^4$Department of Astronomy, Columbia University, Mail Code 5246, 550 W 120th Street, New York, NY 10027, USA }

\begin{abstract}
We present first results from the Blind Ultra Deep HI Environmental Survey (BUDHIES)  of the Westerbork Synthesis Radio Telescope (WSRT).  Our survey is the first direct imaging study of neutral atomic hydrogen gas in galaxies at a redshift where evolutionary processes begin to show.  
In this letter we investigate star formation, HI-content, and galaxy morphology, as a function of environment in Abell 2192 (at $z=0.1876$). 
Using  a 3-dimensional visualization technique, we find that Abell 2192 is 
a cluster in the process of forming, with significant substructure in it. 
We distinguish 4 structures that are separated in  redshift and/or space. 
The richest structure is the baby cluster itself, with a  core of elliptical galaxies that coincides with (weak) X-ray emission, almost no HI-detections, and suppressed star formation. 
Surrounding the cluster, we find a compact group where galaxies pre-process before falling into the cluster, and a scattered population of  ``field-like'' galaxies showing more star formation and HI-detections. 
This cluster proves to be an excellent laboratory to understand the fate of the HI gas in the framework of galaxy evolution. 
We clearly see that the HI gas and the star formation correlate  with morphology and environment at $z\sim0.2$. In particular, the fraction of HI-detections is significantly affected by the environment. The effect starts to kick in in low mass groups that pre-process the galaxies before they enter the cluster. 
Our results suggest that by the time the group galaxies fall into the cluster, they are  already devoid of HI.
\end{abstract}


\section{Introduction}

It has been established that the shaping of galaxies, as well as their star formation activity, strongly depend on their environment \citep[e.g.][]{Dressler1980}. Recent large surveys \citep[SDSS, 2dF,][]{lewis02,gomez03} and studies of galaxy groups, cluster outskirts and filaments \citep[e.g.][]{Treu2003, Wilman2009, Fadda2008, Porter2008} have further shown that the environmental dependencies extend to lower density environments,  and have suggested that galaxies may be ``preprocessed'' before they fall into clusters. 
Clusters of galaxies, in relation to the large scale structure in which they are embedded, thus offer
a unique laboratory to study the effects of  environments on the properties of their
constituent galaxies. 

The evolution of galaxy properties in clusters is strong even in the last few Gyrs: the
fraction of blue galaxies was higher in the past \citep[Butcher-Oemler B-O effect,][]{bo78}, and the relative number of S0 galaxies increases with time, at the expense of the spiral population \citep{Dressler1997,fasano2000}.
Similar trends are also found in the 
field \citep[]{Bell2007, Oesch2010}, where the relative number of red, passively evolving, early-type galaxies increases with time.

Much work has been done to understand the physical processes driving galaxy evolution, at low and high redshift, and at many wavelengths. In particular, HI in and around galaxies is a very useful tool to understand galaxy formation and evolution, as it is the key ingredient for forming stars, and a sensitive tracer of environmental processes. 
Observations  \citep[e.g.][]{Cayatte1990,BravoAlfaro2000,BravoAlfaro2001,PoggiantiVG2001,Kenney2004,crowl05a,Chung2007,Chung2009,Abramson2011,Scott2010,Scott2012} have shown with impressive amount of detail, that the HI gas in galaxies is disturbed and eventually truncated and exhausted in clusters, and simulations \citep[see][for a review]{Roediger2009} have suggested that this happens via ram pressure stripping and gravitational interactions \citep[e.g. ][]{Vollmer2003,TonnesenBryan2009,Kapferer2009}. However, due to technological limitations, studies of the HI content in galaxies have so far mostly been carried out in the local universe.

To address the question where, how and why star-forming spiral galaxies get transformed
into passive early-type galaxies, we have embarked on a Blind Ultra Deep HI Environmental Survey (BUDHIES) with the  
Westerbork Synthesis Radio telescope (WSRT). We refer to \citet[][]{Verheijen2007} for technical details on the HI survey. The strategy has been to study in detail two galaxy clusters, at z$\simeq$0.2 and the large scale structure in which they are embedded.  
The unique aspect of our study is that, for the first time, we have accurate measurements of the HI content in galaxies in different environments  at intermediate redshift. 

Our study is the first where optical properties and gas content are combined at a
redshift where evolutionary effects begin to show, and in a volume large enough to sample all environments, ranging from voids to cluster cores.

The surveyed clusters, Abell 2192 and 963 (A2192 and A963 from now on), are very distinct. A2192, at z=0.188, is less massive and more diffuse than A963. 
%
In this Letter we present first results on the effect of environment on the incidence of star-forming and HI-detected galaxies in A2192. The complete analysis of the two clusters and the large scale structure around them will be presented in a series of papers. 

\section[]{DATA}
\label{sec:data}

BUDHIES is a deep HI survey of galaxies in two clusters at $0.16\leq z\leq0.22$ and the large scale structure around them, with an effective volume depth of 328 Mpc and a coverage on the sky of $\sim$12$\times$12 Mpc for each cluster, that allow us to study the large scale structure around the two Abell clusters. This campaign was possible thanks to a new  back-end and the low system temperature of the WSRT. 
The results of a pilot study are presented in \citet[][]{Verheijen2007}. The complete survey detected 118 galaxies in A963 and 42 in A2192, with HI mass  $M_{HI}\gtrsim2\times10^9\rm M_{\sun}$ (5 sigma) assuming
a typical width of 100 km/s. 

Additionally, we have obtained $B$- and $R$-band imaging with the Isaac Newton Telescope, from which we obtained photometry and morphologies. 
We visually classified the galaxies into early (elliptical and S0) and late-types (spiral, irregular, or interacting).  
We also computed stellar masses from the available SDSS photometry, using the method described in \citet{Zibetti2009}, and a Kroupa initial mass function (see Jaff\'e et al.,~in prep.) 
Furthermore, we acquired deep NUV and FUV imaging with GALEX (Montero-Casta\~no et al. in prep.), 
as well as other auxiliary data that will be presented in future papers.  

A few redshifts are available from the literature and the HI observations in the field of A2192 (15 from SDSS, 44 spectra taken with WIYN/Hydra in the redshift range of the HI observations, and 42 from WSRT).

In order to study the clusters and the large scale structure around them, we obtained additional optical spectroscopy for galaxies in and around our clusters using AutoFib2+WYFFOS (AF2), the multi-object, wide-field, fiber spectrograph mounted at the William Herschel Telescope (WHT). The spectroscopy is thoroughly described in Jaff\'e et al. (in prep.). 
In short, we targeted galaxies with 
colours consistent with the red-sequence and blue cloud at the cluster redshift, down to  $R=19.4$  ($R=20.4$ for HI-detected galaxies). 
The optical AF2/WYFFOS WHT spectroscopy was reduced with the new AF2 data-reduction pipeline. 

The AF2/WHT spectroscopy yielded 377 new redshifts in A2192. Redshifts were  measured from emission lines when possible (mostly [OII]) and from absorption lines (typically Ca H\&K).

We also measured rest-frame equivalent widths (EW) of the [OII]3727$\rm\AA{}$ line,  
which we used as a proxy for ongoing star formation: we considered galaxies as ``star-forming'' if EW[OII] $\geq4\rm\AA{}$ in emission.
We applied the same method to the WIYN spectra. For galaxies with only SDSS spectra, we used EW[OII]$=$EW[OII]3726$+$EW[OII]3729 from DR7.

\section{RESULTS}
\label{sec:results}

\subsection{Defining environment}
\label{envsec}

\begin{figure}
\begin{center}
  \includegraphics[width=0.49\textwidth]{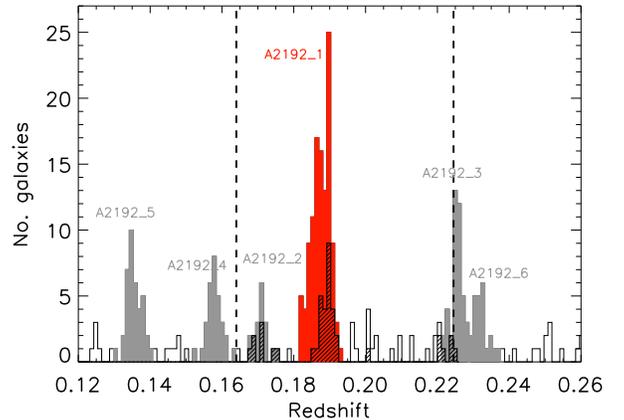}
\end{center} 
\caption{The redshift distribution for galaxies in A2192. Redshifts are mainly taken from the WHT observations, although we also include redshifts from SDSS, WIYN, and WSRT observations. Only secure redshifts were used to construct this histogram and characterize the structures. The focus of this Letter is the main cluster at $z=0.1876$ (A2192\_1, red histogram). However, additional groups of galaxies are also identified in the plot in filled gray histograms for future reference. 
The vertical dashed lines delimit the redshift range of the HI observations, and the filled-dashed histogram shows the distribution of the HI-detected galaxies.}
 \label{zhist}
\end{figure}

\begin{figure}
\begin{center}
  \includegraphics[width=0.49\textwidth]{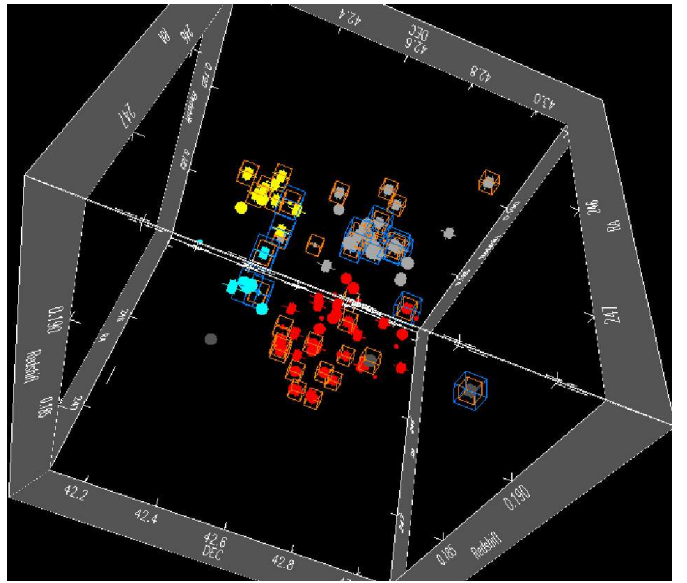}
\end{center} 
\caption{3D ($\alpha$-$\delta$-redshift) visualization of galaxies in A2192\_1 \citep[made with the S2PLOT progamming library, ][]{Barnes2006}.  The galaxies are color-coded according to the substructure they belong to:  Red correspond to A2192\_1a, light gray to A2192\_1b, turquoise to A2192\_1c, and  yellow to A2192\_1d, as in  Figure~\ref{zhist2} (darker gray represents galaxies not associated to any of the mentioned structures).  HI-detected galaxies are surrounded by blue open cubes, while star-forming galaxies are surrounded by (smaller) open orange cubes. Morphology is illustrated with different symbols: big spheres correspond to early-type galaxies, solid cubes with spikes are late-types, and small symbols are galaxies with undetermined morphology. Only galaxies in the ``reduced sample'' are used.  The $\alpha$-$\delta$ plane is $\cong$14$\times$14 Mpc. The white segment indicates 1Mpc at the cluster redshift. An animated version of this plot is available in the electronic version of the paper, or alternativelly, at: http://www.nottingham.ac.uk/$\sim$ppxyj/Jaffe\_ApJL\_2012\_Fig2\_movie.mpeg.}
 \label{3d_plot}
\end{figure}

Figure~\ref{zhist} shows the redshift distribution of the galaxies in A2192. 
There is one main structure centered at $z=0.1876$ (A2192\_1, red solid histogram) with a large number of galaxies (101), and several foreground and background galaxy clusters/groups (gray hystograms) that will be fully studied in Jaff\'e et al.~(in prep). 
In this letter, we focus on A2192\_1. 

The cluster velocity dispersion ($\sigma_{\rm cl}$) and central redshift  ($z_{\rm c}$) were computed following \citet*{bfg90} and are listed in Table~\ref{structures}. We considered cluster members those galaxies with radial velocities within $3\sigma$ from $z_{\rm c}$. From $\sigma_{\rm cl}$, we also calculated $R_{200}$, the radius delimiting a sphere that has mean enclosed density equal to 200 times the critical density. 

\begin{figure}
\begin{center}
  \includegraphics[width=0.49\textwidth]{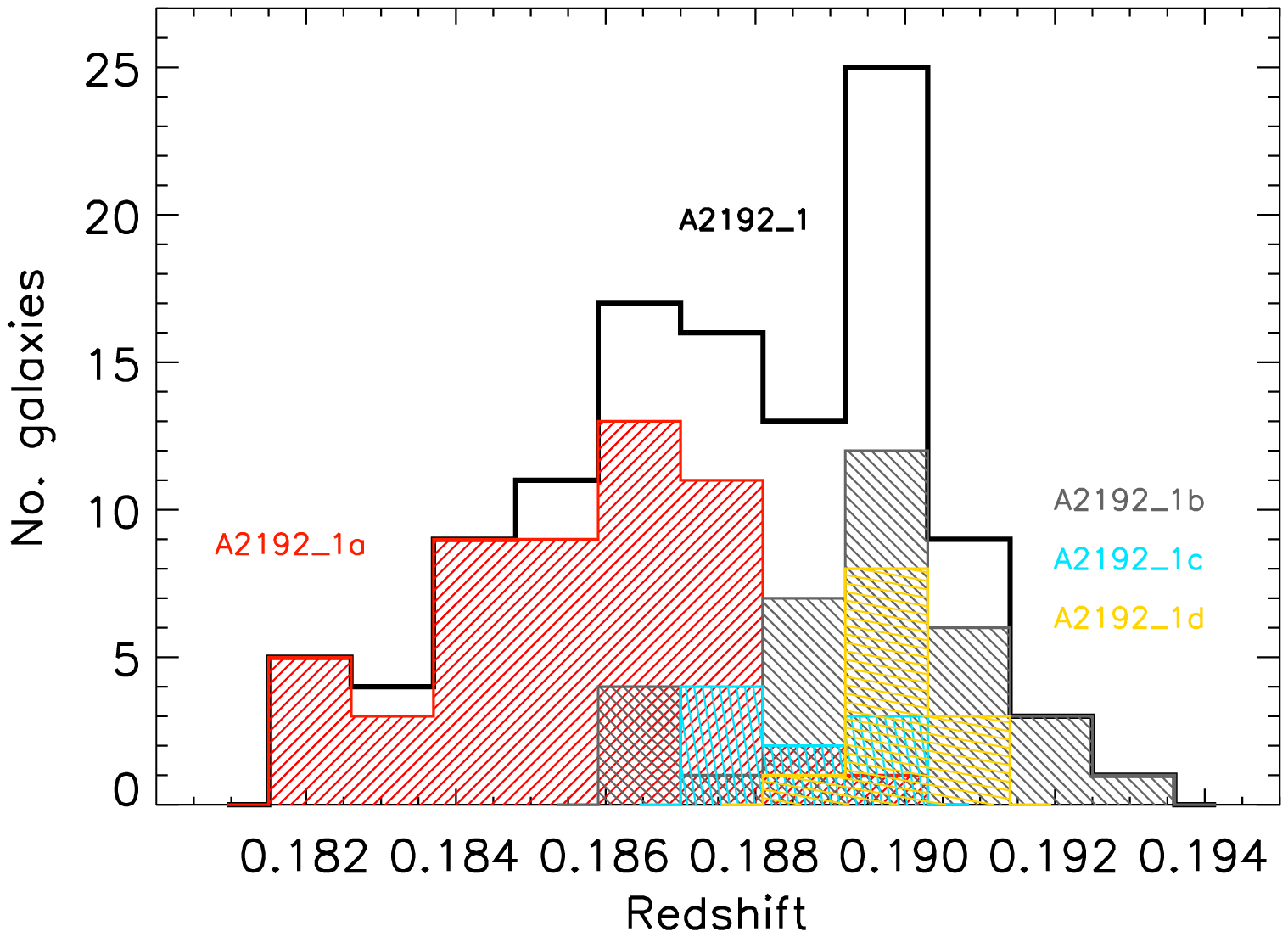}
  \includegraphics[width=0.49\textwidth]{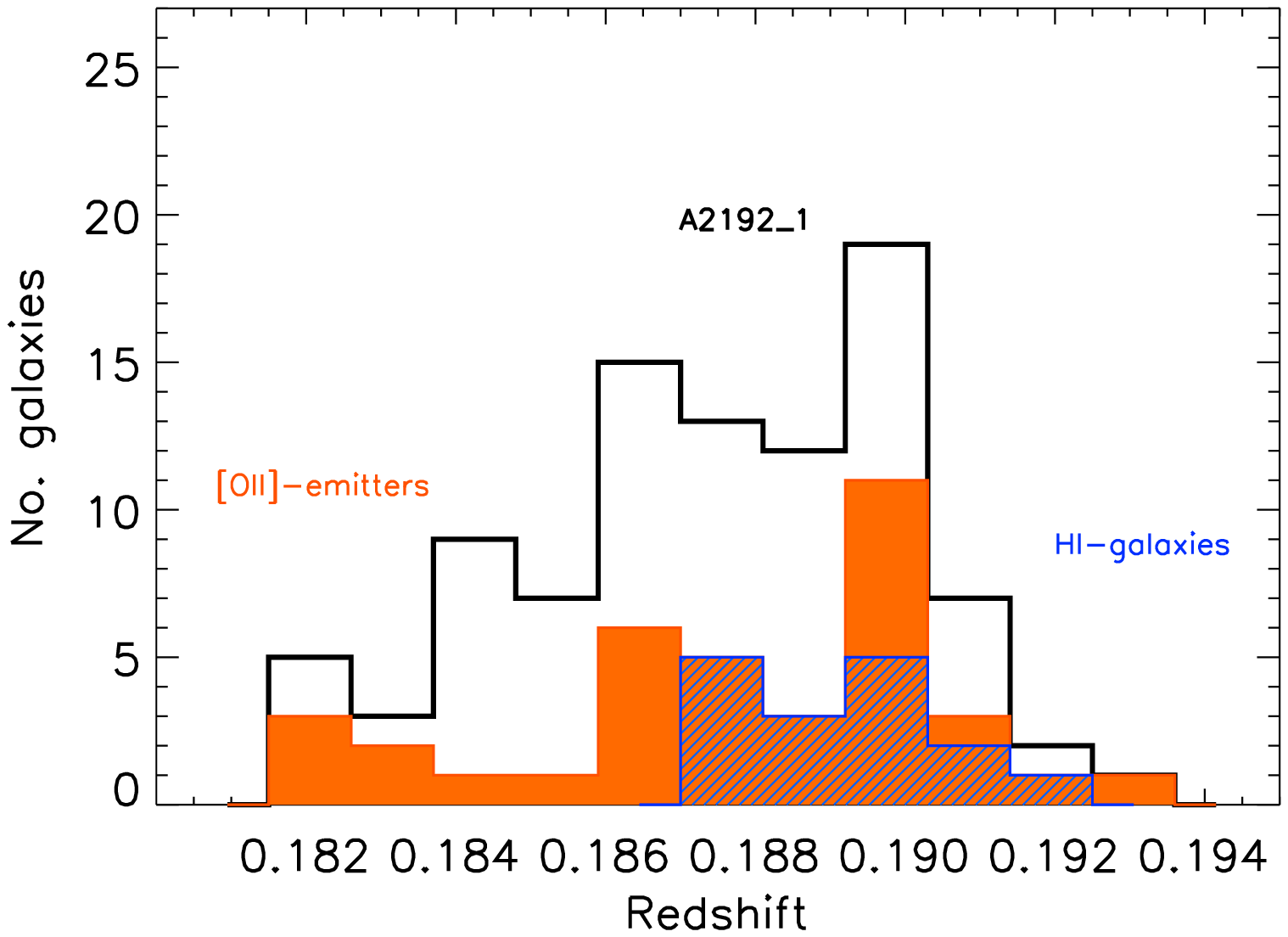}
\end{center} 
\caption{\textit{Top:} The overall redshift distribution of the galaxies in A2192\_1 (open black histogram), and the four substructures found within A2192\_1 in colored filled-dashed histograms, (same colors as in Figure~\ref{3d_plot}).
 \textit{Bottom: } The redshift distribution of A2192\_1 galaxies is shown, in addition to the distribution of the HI-detected galaxies (blue filled-dashed histogram ), and the galaxies with [OII] emission (solid dark orange histogram). In the top plot we use all galaxies in A2192\_1, and in the bottom plot we use the ``reduced sample''.}
 \label{zhist2}
\end{figure}

From the histogram of Figure~\ref{zhist}, it is hinted that A2192\_1's redshift distribution  is not entirely Gaussian. This is shown more clearly on the top panel of Figure~\ref{zhist2}, where only the galaxies in A2192\_1 are plotted (open black histogram). The distribution shows a double-peak shape, strongly suggesting that this ``cluster'' is not virialized, and contains significant substructure.

To identify the substructures inside A2192\_1, we combined the redshift information with the distribution of galaxies on the sky. 
This yielded a 3-dimensional (3D) picture, in which it becomes evident that the cluster has indeed 4 distinct substructures.  
These are illustrated in Figure~\ref{3d_plot}, where the different substructures are represented with different colors. 
Figure~\ref{3d_plot} shows a reduced number of galaxies as a consequence of limiting the plotted sample to galaxies with optical spectroscopy (and thus EW[OII] measurements), and keeping our spectroscopic magnitude limit ($R\leqslant19.4$ mag, i.e. excluding the faintest HI galaxies). Such ``reduced sample'', is defined to compare the fraction of star-forming galaxies with the fraction of HI-galaxies within the structures (see Section~\ref{HIsec}). 
Substructures within A2192\_1 were identified on the basis of all galaxies with a reliable redshift, although not all these galaxies are shown in  Figure~\ref{3d_plot} for clarity. 
We note that we identified these structures without using any information on galaxy morphology, HI-content or [OII]-emission. We identified them only by carefully inspecting the 3D distribution of the galaxies searching for galaxy groups separated in redshift and/or space. 

The different substructures are also presented in the redshift histogram at the top of Figure~\ref{zhist2}.
We named these substructures A2192\_1a (red in Figure~\ref{3d_plot} and the upper panel of Figure~\ref{zhist2}), A2192\_1b (gray), A2192\_1c (turquoise),  and A2192\_1d (yellow). The properties of the structures are listed in Table~\ref{structures} and discussed in Section~\ref{HIsec}.

\subsection{HI, star formation, morphology, and stellar mass as a function of environment}
\label{HIsec}

Figure~\ref{3d_plot} shows detections of HI and star formation activity, as well as morphology of the galaxies in each of the structures in A2192\_1. 
The bottom panel of Figure~\ref{zhist2} further shows the redshift distribution of the galaxies in A2192\_1 (black open histogram), with additional histograms for HI-detected (blue) and star-forming (orange) galaxies. 
Interestingly, the HI detections are offset to higher redshift with respect to A2192\_1a. 
Star forming galaxies, on the other hand, are spread across all redshifts.

In the following we analyze the characteristics of each substructure and its  galaxy population, as presented in Figure~\ref{3d_plot}, to unveil the effect of environment on the galaxies' properties. 

\begin{itemize}
 \item \textbf{A2192\_1a:} This structure is the main cluster. It is well defined in space,  is the richest and most massive of all substructures, and has a nearly Gaussian redshift distribution, suggesting it is gravitationally-bound and virialized. It is an intermediate-mass cluster ($2.27\times10^{14}M_{\sun}$, or $\sigma_{\rm cl}=530$ km/s), and  is clearly still assembling. 
 Additional evidence that supports the presence of a cluster comes from  the X-ray emission in A2192, that, although weak, coincides with the group of early-type galaxies at the core of A2192\_1a. Moreover, the derived virial mass is consistent with the X-ray luminosity \citep[$L_{X}\simeq7\times10^{43}h_{100}^{-2}$ erg/s; ][]{Voges99}.

A2192\_1a is the least HI-rich of the 4 structures. Only 1 out of 46 (2\%) of the galaxies in this cluster are HI-detections, and these are located at a projected distance of $\sim2\times R_{200}$ of the center, at the boundary of the cluster, where A2192\_1a meets A2192\_1b. Nonetheless, A2192\_1a has a significant fraction (28\%) of galaxies with [OII] emission. All of the HI-detected galaxies, as well as the majority of the star-forming galaxies are late types, as one would expect. 
Overall, the galaxies in this cluster, although  devoid of HI, still host star formation, especially towards the outskirts.

 \item\textbf {A2192\_1b:} Although very rich, it is very dispersed in space. This ``structure'' is unlikely to be a bound group, as its shape and location are consistent with that of a population of field galaxies.
The late-type galaxy population in this ensemble of ``field'' galaxies, has 40\% of HI-detections and 80\% of emission-line galaxies. All the HI-detected galaxies have [OII] emission, and all HI and [OII] galaxies were classified as late-type galaxies, as one could expect. 
However, it is interesting to note the presence of a few (8 out of 25) early-type galaxies with no HI nor [OII] emission. It is possible that some of the early-types in this group are truly field E/S0 galaxies, but it is also possible that some are early spirals that were classified as E/S0. 

 \item \textbf{A2192\_1c:} With less than 10 galaxies, and considerably spread in space, it is unlikely to be a gravitationally bound system. 
 Although spatially separated from the other structures, it resembles the ``field''  (like A2192\_1b). 
 Its galaxies span the whole morphology range, with a few more late-type galaxies than early-types. 
The fraction of late-type galaxies detected in HI is similar to that in A2192\_1b.

 \item \textbf{A2192\_1d:} Likely to be a galaxy group, it is very compact and clearly separated in space.
 Its dynamical mass is small ($6.34\times10^{12} M_{\sun}$, based on $\sigma_{\rm cl}=161$ km/s), and it is composed mainly of late-type galaxies. 63\% of the late-type galaxies are star-forming, and a few (between 13 and 26\%) have HI reservoirs. All the galaxies with HI or [OII] emission are late-types, while the rest are early-types. As in A2192\_1, the HI-detected galaxies are in the outskirts of the group (at $>1\times R_{200}$), with the exception of one. The star formation on the other hand is more widely spread. The properties of this group places it somewhere in between the cluster (A2192\_1a) and the field (A2192\_1b), which suggests that it might be a place where galaxies are being ``pre-processed''. It is also possible that the gas gets removed from group galaxies as the groups become more compact \citep{VerdesMontenegro2001}, or that the gas in this compact group gets removed as it moves closer to the cluster, or  both.

\end{itemize}

The structures surrounding the main cluster (A2192\_1a) are all at a distance well within A2192\_1a's turnover radius \citep[c.f.][]{RinesDiaferio2006}, implying that they will eventually be accreted by the cluster.

\begin{figure}
\begin{center}
  \includegraphics[width=0.49\textwidth]{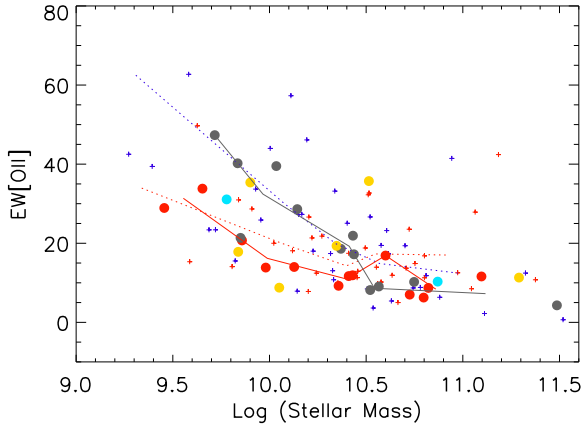}
\end{center} 
\caption{EW[OII] vs. stellar mass for the different structures in A2192\_1 (filled circles, same colors as in Figure~\ref{3d_plot}). For comparison, we also include the field population at a similar redshift (in A2192 and A963; blue crosses), and other clusters in the redshift range (same as in gray histograms of Figure~\ref{zhist}; red crosses). Lines correspond to average EW[OII] in several mass bins for field galaxies (dotted blue), A2192\_1a (solid red), A2192\_1b (solid gray), and other clusters (dotted red). We adopt the same ``reduced sample'' of Figure~\ref{3d_plot}.}
 \label{ew_mass}
\end{figure}

To assess whether the star formation rate in star-forming galaxies is reduced/suppressed by the environment, we compared the EW[OII] in each structure at fixed galaxy stellar mass. As Figure~\ref{ew_mass} shows, star-forming galaxies in A2192\_1a have a suppressed star formation (i.e. lower EW[OII]) at stellar masses $\lesssim 3 \times 10^{10}$, while A2192\_1b is consistent with the field population. Unfortunately, the numbers are too low to make conclusions about A2192\_1c and A2192\_1d.  
Nonetheless, (near and far) UV colour-colour plots reveal a clear separation of the structures (see Montero-Casta\~no et al. in prep.):  
A2192\_1a and A2192\_1d  have redder (older) galaxies, when compared with A2192\_1b and A2192\_1c, supporting our environmental definitions.

\section{SUMMARY}
\label{sec:concl}

We have studied the fraction of star-forming and HI-detected galaxies in A2192. 
In Deshev et al (in prep.) we present the ultra-deep HI-survey covering $0.16\leq z \leq0.22$, which is the core of this study.
In this Letter we present first results from combining the HI-observations with optical spectroscopy and imaging. 
The redshift distribution reveals several structures. 
We focus on the central overdensity at redshift $z=0.1876$ (A2192\_1). 
This ``cluster'' shows clear signs of substructure, both in its redshift and spatial distribution. 
By inspecting it in 3D ($\alpha$-$\delta$-z) we identify 4 distinct substructures:  
a medium-sized galaxy cluster in the process of forming, 
surrounded by a less massive infalling group, 
and a spread population of ``field-like'' galaxies. 

Further evidence from observed galaxy properties support the idea that the cluster is still assembling.   
Most notably, the main cluster (A2192\_1a) is practically devoid of galaxies with $M_{HI}\geq2\times10^{9}M_{\odot}$, has suppressed star formation, and a core of red early-type galaxies that coincides with the X-ray center. The cluster is surrounded by scattered ``field'' (A2192\_1b and A2192\_1c) of mostly late-type galaxies that are actively star-forming and still preserve their HI reservoirs. Moreover, there is a spatially separated compact group (A2192\_1d), where galaxies show signs of being pre-processed before the group falls into A2192\_1a. 
This suggests that the environmental effect starts to kick in in low-mass groups that pre-process the galaxies before they enter the cluster. In the case of cluster growth via group accretion, it is likely that by the time the group galaxies fall into the cluster, they are  already devoid of HI.

A2192 proves to be an excellent laboratory to understand the fate of the HI gas in the framework of galaxy evolution. In particular, we clearly see that the HI gas, as well as the star formation in late type galaxies correlates with environment at $z\sim0.2$. Our results suggest that the HI gas gets removed and star formation suppressed progressively,  
from the lowest mass galaxy groups, to cluster-sized structures, as smaller structures get assembled into larger structures.

\section*{Acknowledgements}

YLJ and BMP acknowledge financial support from ASI  contract I/099/10/0. 
YLJ thanks Richard Jackson and Ian Skillen for their support on the new AF2 pipeline, and Daniela Bettoni for useful discussions. 
This work was supported in part by the National Science Foundation under grant No. 1009476 to Columbia University. 
We are grateful for support from a Da Vinci Professorship at the Kapteyn Institute.

\begin{table*}
\begin{center}
 \caption{Characteristics of A2192\_1, and its composing substructures (see Figures~\ref{zhist2}, and~\ref{3d_plot}). The columns are: Name of cluster/structure;  Median redshift ($z_{\rm c}$); Number of galaxies ($R\leq19.4$);  Fraction of HI-detected galaxies;   Fraction of galaxies with EW([OII]$\geq4\rm\AA{}$;  Fraction of late-type galaxies with HI emission;  Fraction of late-type galaxies with [OII] emission;  Cluster/group velocity dispersion;  R$_{200}$ (``--'' is placed when it is not applicable);  Comments. The numbers inside parenthesis at the bottom of the quoted values correspond to the ``reduced sample'' we used to calculate HI and OII fractions of columns 4-7. Error on the fractions correspond to the confidence intervals (c$\approx$0.683) for binomial populations, from a beta distribution \citep[see][]{Cameron2011}.} 
\label{structures}
\begin{tabular}{lccccccccl} 
\hline\\[-1mm]
Name of     	&  $z_{\rm c}$   	& No. 		&Frac. HI-  		&Frac. [OII]		& Frac. LTGs		& Frac. LTGs		& $\sigma_{\rm cl}$ 	& R$_{200}$ 	& Remarks \\ 
structure  	&     		 	& members	&galaxies 		&emitters		& with HI		& with [OII]		&  (km$/$s) 	    	& (Mpc) 	&(and color in Figures) \\
\hline\\[-2mm]
A2192\_1	& 0.1876		& 101		&  14$^{+4}_{-3}$\% 	& 39$\pm5$\%		&21$^{+7}_{-5}$\%	&56$\pm7$\%		& 653$\pm$62		& 1.43		& Central ``cluster'' in A2192 \\ 
 		& 			& (93)		& (13/93)		&(36/93)		&(10/48)		&(27/48)		&			&		&  (red histogram in Fig.~\ref{zhist}) \\
\hline\\[-2mm]
A2192\_1a$^{*}$	& 0.1859		& 50		& 2$^{+5}_{-1}$\%	& 28$^{+7}_{-6}$\%	&5$^{+9}_{-1}$\%	&36$^{+11}_{-9}$\%	& 530$\pm$56		&1.16	 	& Richest \& more massive  \\

		& 			& (46)		& (1/46)		&(13/46)		& (1/22)		&(8/22)			&			& 		& substructure in A2192\_1 (red  \\
		& 			& 		&       		&        		&       		& 			&			& 		& in top panel of Fig.~\ref{zhist2} and~\ref{3d_plot}) \\
A2192\_1b	& 0.1898		& 29		&  24$^{+10}_{-6}$\%	& 52$^{+9}_{-10}$\%	&40$^{+13}_{-11}$\%	&80$^{+7}_{-14}$\%	&	--		& 	--	& Disperse, ``field-like'' \\
		& 			& (25)		& (6/25)		&(13/25)		&(6/15)			&(12/15)		&			& 		& (gray) \\
A2192\_1c$^{\dag}$	& 0.1881 	& 8		& 38$^{+18}_{-13}$\%	& 25$^{+19}_{-9}$\%	&50$\pm25$\%		&50$\pm25$\%&	--		&	--	& Poor, disperse \\
		& 			& (8)		& (3/8)			&(2/8)			&(1/2)			&(1/2)			&			&		& (turquoise)	\\
A2192\_1d$^{* \dag}$	& 0.1902	& 11		& 18$^{+16}_{-6}$\%	& 55$^{+13}_{-15}$\%	&13$^{+20}_{-5}$\%	&63$^{+13}_{-18}$\%	&161$\pm52^{*}$	&0.35		& Small, compact group\\
		& 			& (11)		& (2/11)		&(6/11)			&(1/8)			&(5/8)			&*$\sigma_{\rm Gapper}$	&		& (yellow)	\\
\hline\\[-2mm]

\end{tabular} 

*The luminosity-weighted geometric centers of A2192\_1a and A2192\_1d are at  $\alpha=246.64774^{\circ}, \delta=+42.727723^{\circ}$, and  $\alpha=246.48559^{\circ}, \delta=+42.380932^{\circ}$ respectivelly.\\
$\dag$ Because these structures are located far from the pointing center of WSRT's primary beam, the HI fractions 
 suffer from beam attenuation effects (fully treated in Deshev et al. in prep.). We estimate that the fraction of HI galaxies among the late/types can at most increase by a factor of 2.\\
\end{center}
\end{table*}

\setlength{\bibhang}{2.0em}

\end{document}